\documentclass[12pt,psfig]{article}
\usepackage{amssymb}
\usepackage{amsmath,amsthm}
\usepackage{amsfonts}
\usepackage{graphicx}
\usepackage{graphics}
\usepackage{indentfirst}
\numberwithin{equation}{section}

\usepackage{hyperref}

\begin{document}
\begin{center}\Large\textbf{Left-Right
Entanglement Entropy for a D$p$-brane 
with Dynamics and Background Fields}
\end{center}
\vspace{0.75cm}
\begin{center}{\large Shirin Teymourtashlou and \large Davoud
Kamani}\end{center}
\begin{center}
\textsl{\small{Department of Physics, Amirkabir University of
Technology (Tehran Polytechnic), Iran \\
P.O.Box: 15875-4413 \\
e-mails: sh.teymourtash@aut.ac.ir , kamani@aut.ac.ir \\}}
\end{center}
\vspace{0.5cm}

\begin{abstract}

We investigate the left-right entanglement entropy of
a boundary state, corresponding to a dynamical
D$p$-brane with the internal and background fields.
We assume that the brane has a tangential
linear motion and a rotation, and is dressed
with an internal $U(1)$ gauge potential and
the Kalb-Ramond tensor field $B_{\mu\nu}$.
We derive the entanglement entropy
via the R\'{e}nyi entropy by applying the replica trick.
Our calculations will be in the context of the bosonic
string theory.

\end{abstract}



\newpage
\section{Introduction}

In a composite quantum system, 
which consists of subsystems, 
entanglement relates the different parts of the system.
The subsystems can become 
entangled if the quantum state of each subsystem
cannot be described independent of the states of the
other subsystems. In fact, the 
quantum systems are capable to 
become entangled through the various 
types of processes such as 
interactions, particles creation and etc. 
For instance, in the decay of the subatomic 
particles, because of the conservation laws, the measured 
quantum labels for the daughter 
particles are highly correlated.
Traditionally, for quantifying entanglement,
geometric setups have been intensively studied in the
literature \cite{1}-\cite{6}. 
Entanglement entropy is a favorable quantity
for measuring the entanglement between the subsystems. Also,
this quantity has been drastically studied
in the context of the AdS/CFT \cite{7, 8}.

At first consider a bipartite system with the subsystems
A and B. The division can 
occur in the Hilbert space (instead of the configuration
space), i.e.
$\mathcal{H}=\mathcal{H}_{\rm A}\otimes \mathcal{H}_{\rm B}$.
Let $|a_i\rangle$ and $|b_j\rangle$ be
the eigen-bases which span  
$\mathcal{H}_{\rm A}$ and $\mathcal{H}_{\rm B}$,
respectively. Thus, $|a_i\rangle \otimes |b_j\rangle$
forms an eigen-basis for $\mathcal{H}$. Hence,
a generic state $|\psi\rangle$ in $\mathcal{H}$
possesses the expansion 
\begin{equation}
|\psi\rangle=\sum_{i,j} c_{ij}
|a_i\rangle \otimes |b_j\rangle.
\end{equation}
If the correlation coefficients $c_{ij}$ can be decomposed,
e.g., as
$c_{ij}=\alpha_i \beta_j$ we acquire the product state  
$|\psi\rangle =|\psi_{\rm A}\rangle 
\otimes |\psi_{\rm B}\rangle$.
In this case the subsystems A and B are not 
entangled. For the case 
$|\psi\rangle \neq|\psi_{\rm A}\rangle 
\otimes |\psi_{\rm B}\rangle$
we have an entangled system.

In our system the left- and right-moving oscillating
modes of closed strings are the bases of the two subsystems,
hence, the Hilbert space possesses the factorized form
$\mathcal{H}=\mathcal{H}_{\rm L}\otimes
\mathcal{H}_{\rm R}$. The Schmidt decomposition of 
the boundary state with respect to the 
left- and right-moving modes can be written as
\cite{9, 10},
\begin{equation}
|B\rangle=\mathcal{N} \sum_{\overrightarrow{m}} 
|\overrightarrow{m}\rangle \otimes 
|U\tilde{\overrightarrow{m}}\rangle,
\end{equation}
where the states $|\overrightarrow{m}\rangle$ and 
$|\tilde{\overrightarrow{m}}\rangle$ are complete 
orthonormal bases for $\mathcal{H}_{\rm L}$ and 
$\mathcal{H}_{\rm R}$, 
and $U$ is an anti-unitary operator which acts
on $\mathcal{H}_{\rm R}$. 
Both states $|\overrightarrow{m}\rangle$ and 
$|\tilde{\overrightarrow{m}}\rangle$ 
depend on a set of the integer numbers 
$\{m_1,m_2, \cdots\}$. Now
consider Eq. (1.1) for the maximally entangled case, i.e. 
$c_{ij}=c \delta_{ij}$, and compare it
with Eq. (1.2). This comparing clarifies that 
the decomposition (1.2) represents the boundary state $|B\rangle$ 
as a maximally entangled state of the left- and right-moving 
modes. Thus, 
we can choose the boundary state as our composite system 
and the left- and right-moving modes of closed strings as 
its subsystems.

On the other hand, the D-branes as dynamical
objects are essential for studying different areas
of string theory. We shall investigate one of the
attractive characteristics of a D-brane, i.e. the so called
left-right entanglement entropy (LREE) \cite{11}-\cite{14}.
The left-right entanglement is a
non-geometrical version of the entanglement.
Since the boundary state accurately encodes all properties of
a D-brane \cite{15}-\cite{20}, it is a useful tool
for investigating the LREE corresponding to the D-brane.

Zayas and Quiroz previously worked out
the LREE for a one-dimensional boundary state in a free
bosonic 2D CFT with
the Dirichlet or the Neumann boundary condition \cite{11}.
Besides, they derived the LREE for the
bare-static D-branes \cite{13}.
By making use of their approach, in this paper
we shall obtain the LREE
for a bosonic D$p$-brane which is dressed by the
Kalb-Ramond background field $B_{\mu\nu}$ and
an internal $U(1)$ gauge potential $A_\alpha$
which lives in the brane worldvolume. In addition,
we impose a tangential dynamics to the brane,
which includes linear motion and rotation.
We shall observe that the LREE of our setup may be
interpreted as a thermodynamical entropy. 

In fact, the entanglement entropy of the D-branes
potentially has relation with the black holes
entropies \cite{6, 21}. Therefore, we are
motivated to investigate the LREE of a special
D$p$-brane. Precisely, a brane configuration
with the background and internal
fields can be corresponded to a charged black hole.
Besides, a dynamical brane, especially those with internal
rotations, may be associated with a
rotating black hole. Ultimately, the LREE
of our brane configuration may find a connection with the
entropy of the charged-rotating black holes.

Since the extracted quantities of the bosonic string theory 
are similar to their counterparts in 
the NS-NS sector of the superstring
theory we shall begin our calculations for the foregoing bosonic
D-brane. Beside, the bosonic computations are more simple
than the superstring calculations. 
Hopefully, in the subsequent works we shall extend 
our calculations to the supersymmetric version.

The paper is organized as follows. In Sec. 2, we shall
introduce the boundary state, corresponding to the
D$p$-brane, then, the interaction amplitude between
two parallel and identical D$p$-branes will be
introduced. This amplitude is
required for calculating the R\'{e}nyi entropy.
In Sec. 3, we shall compute the LREE for
a bare-static D$p$-brane and for a dressed-dynamical one.
We shall terminate this section with a thermodynamical
interpretation of the LREE of our system.
In Sec. 4, some simple examples will be presented to
clarify the parametric dependence of the setup.
Section 5 is devoted to the conclusions.

\section{The dressed-dynamical D$p$-branes:
boundary state and interaction}

\subsection{The boundary state}

In the beginning we introduce the boundary state, associated 
with a D$p$-brane with tangential dynamics,
in the presence of the antisymmetric tensor
$B_{\mu\nu}$ and the internal gauge field $A_{\alpha}$. Thus,
we apply the following closed string action
\begin{eqnarray}
S=&-&\frac{1}{4\pi\alpha'}\int_{\Sigma} d^2\sigma
\left(\sqrt{-h} h^{ab}g_{\mu\nu} \partial_{a}X^\mu
\partial_b X^{\nu} +\varepsilon^{ab}
B_{\mu\nu}\partial_a X^{\mu}
\partial_b X^{\nu}\right)
\nonumber\\
&+&\frac{1}{2\pi\alpha'}\int_{\partial\Sigma}
d\sigma\left(A_{\alpha}
\partial_{\sigma}X^{\alpha}
+\omega_{\alpha\beta}J^{\alpha\beta}_{\tau} \right)~,
\end{eqnarray}
where the indices
$a, b \in \{0,1\}$ are devoted to the string worldsheet
and $\alpha, \beta\in \{0,1,\ldots,p \}$ belong
to the D$p$-brane worldvolume. Let the spacetime
be flat, i.e. $g_{\mu\nu}=\eta_{\mu\nu}$.
In addition, the string worldsheet will be flat.
The tensors $\omega_{\alpha\beta}$ and
$J^{\alpha \beta}_\tau= X^\alpha \partial_\tau
X^\beta -X^\beta \partial_\tau X^\alpha$ indicate
the tangential angular velocity and the
angular momentum density, respectively. The
angular velocity $\omega_{\alpha\beta}$, the
Kalb-Ramond field $B_{\mu \nu}$ and the field
strength of the gauge potential, i.e. $F_{\alpha\beta}$,
are taken to be constant, hence, we utilize the gauge
$A_\alpha =-\frac{1}{2} F_{\alpha\beta} X^\beta$.
Because of the presence of the fields on the brane worldvolume
the Lorentz symmetry breaks down, thus, the
tangential dynamics along the worldvolume directions
obviously is sensible. In the rest of the paper we take
$\alpha'=2$.

The boundary state equations can be
obtained by vanishing of the variation of the action with
respect to $X^\mu$,
\begin{eqnarray}
&~&\left[\left(\eta_{\alpha\beta}+4\omega_{\alpha\beta}\right)
\partial_{\tau}X^{\beta}+\mathcal{F}_{\alpha\beta}
\partial_\sigma X^{\beta}
+ B_{\alpha i}\partial_\sigma X^i\right]_{\tau=0}|B_x\rangle=0 ,
\nonumber\\
&~&\left(X^i-y^i\right)_{\tau=0}|B_x\rangle=0,
\end{eqnarray}
where
$\mathcal{F}_{\alpha\beta} \equiv
B_{\alpha\beta}-F_{\alpha\beta}$.
The Dirichlet directions are shown by
$\{x^i|i = p+1, \ldots,d-1\}$
and the parameters $y^i$ specify the brane position.
One can use the mode expansion of $X^\mu$ to rewrite
the above equations in terms of the closed
string oscillators
\begin{eqnarray}
&~&\left[\left(\eta_{\alpha\beta}+4\omega_{\alpha\beta}
- \mathcal{F}_{\alpha\beta}\right)
\alpha_{m}^{\beta}+
\left(\eta_{\alpha\beta}+4\omega_{\alpha\beta}
+ \mathcal{F}_{\alpha\beta}\right)
\tilde{\alpha}_{-m}^{\beta}
\right]|B_{\rm osc}\rangle=0,
\nonumber\\
&~& \left(\eta_{\alpha\beta}
+4\omega_{\alpha\beta}\right)p^{\beta}
|B\rangle^{(0)}=0\label{eq:2}~,
\end{eqnarray}
for the tangential directions, and
\begin{eqnarray}
&~&(\alpha_{m}^{i}-\tilde{\alpha}_{-m}^{i})
|B_{\rm osc}\rangle=0,
\nonumber\\
&~&(x^i-y^i)|B\rangle^{(0)}=0,
\end{eqnarray}
for the perpendicular directions to the worldvolume.
The following decomposition was also applied
$|B_x\rangle =|B_{\rm osc}\rangle \otimes|B\rangle^{(0)}$.

The second equation of Eq. (\ref{eq:2}) eventuates to
$p^\alpha \det \left(\eta+4\omega\right)=0$.
Thus, there are two possibilities depending on whether
$\left(\eta_{\alpha\beta}+4\omega_{\alpha\beta}\right)$
is invertible or not. We consider the invertible case
which leads to the vanishing tangential momentum
$p^\alpha=0$. Hence, by applying
the commutation relations and the coherent state
formalism we find the zero-mode part and oscillatory
sector of the boundary state as follows
\begin{eqnarray}
|B\rangle^{(0)}&=&\frac{T_p}{2}
\prod^{d-1}_{i=p+1} \delta \left({x}^{i}-y^{i}\right)
|p^i=0\rangle
\prod^{p}_{\alpha =0}
|p^{\alpha}=0\rangle ~
\label{zer},
\end{eqnarray}
\begin{eqnarray}
|B_{\rm osc}\rangle
&=&\sqrt{-\det{M}}\exp\left[{-\sum_{m=1}^{\infty}
\left(\frac{1}{m}\alpha_{-m}^{\mu}S_{\mu\nu}
\tilde{\alpha}_{-m}^{\nu}\right)}\right]
|0\rangle_\alpha \otimes |0\rangle_{\tilde{\alpha}}~,
\label{eq:2.6}
\end{eqnarray}
where $T_p$ is the brane tension,
and the matrix $S_{\mu\nu}$ is defined by
\begin{eqnarray}
S_{\mu\nu}&=&\left(Q_{\alpha \beta} \equiv (M^{-1}
N)_{\alpha\beta},-\delta_{ij}\right)~,
\nonumber\\
M_{\alpha\beta}&=&\eta_{\alpha\beta}+4\omega_{\alpha\beta}
- \mathcal{F}_{\alpha\beta}~,
\nonumber\\
N_{\alpha\beta}&=&\eta_{\alpha\beta}+4\omega_{\alpha\beta}
+\mathcal{F}_{\alpha\beta}~.
\end{eqnarray}
The prefactor in the oscillating part comes from the
normalization of the disk partition function.
For prefactors of the stationary setups see, e.g.,
Ref. \cite{22}. One may define the combination 
$\mathcal{T}_p =T_p\sqrt{-\det{M}}$ 
as an effective tension for the dynamical brane in the 
presence of the internal and background fields.

In fact, the coherent state method enabled us
to acquire the boundary state (\ref{eq:2.6})
under the condition $SS^{\rm T} = {\bf 1}$.
This condition reduces the number of the total
parameters from $3p(p+1)/2$ to $p^2-1$ independent parameters.

In addition to the foregoing sectors of the boundary state,
there also exists a contribution from the conformal ghosts too
\begin{equation}
|B_{\rm gh}\rangle=\exp{\left[\sum_{m=1}^{\infty}
(c_{-m}\tilde{b}_{-m}
-b_{-m} \tilde{c}_{-m})\right]}\frac{c_0+\tilde{c}_0}{2}
|q=1\rangle|\tilde{q}=1\rangle~.
\end{equation}
Therefore, the total bosonic boundary state, corresponding
to the D$p$-brane, is given by
\begin{equation}
|B\rangle =|B_{\rm osc}\rangle \otimes|B\rangle^{(0)}
\otimes|B_{\rm gh}\rangle~.
\end{equation}

\subsection{The amplitude of interaction}

The interaction amplitude of two parallel D$p$-branes
enables us to extract the partition function,
which will be required for computing the LREE.
For calculating the interaction amplitude
we can look at the one-loop
diagram of an open string, stretched between the
branes, or
equivalently study the tree-level diagram of
the exchanged closed string. This equivalence is a
consequence of the conformal invariance of string theory.

Here, we apply the second approach in which the interaction
amplitude is given by the overlap of the
two boundary states, corresponding to the two
dressed-dynamical D$p$-branes,
via the closed string propagator $D$,
\begin{eqnarray}
\mathcal{A} &=& \langle B_1|D|B_2 \rangle~,
\nonumber\\[10pt]
D &=& 4\int_{0}^{\infty}dt~e^{-tH}~,
\end{eqnarray}
where $H$ is the closed string Hamiltonian.
Accordingly, the interaction amplitude finds the feature
\begin{eqnarray}
\mathcal{A}&=&\frac{T_p^2\;V_{p+1}}{8(2\pi)^{d-p-1}}
\sqrt{\det(M^{\rm T}_1 M_2)}
\int_{0}^{\infty}dt\bigg{[}e^{(d-2)\pi t/6}
\nonumber\\[10pt]
&\times& \left(\sqrt{\frac{1}{2t}}\right)^{d-p-1}
\exp\left( {-\frac{1}{8\pi t}
\sum_{i=p+1}^{d-1}{\left(y_{1}^{i}-y_{2}^{i}\right)^2}} \right)
~\nonumber\\[10pt]
&\times & \prod_{n=1}^\infty \bigg{(}
\det[\mathbf{1}-Q^{\rm T}_1 Q_2 e^{-4n\pi t}]^{-1}~
\left(1- e^{-4n\pi t}\right)^{p-d+3}\bigg{)}\bigg{]} 
\label{eq:2.11},
\end{eqnarray}
where $V_{p+1}$ is the brane worldvolume. The first 
exponential comes from the zero-point energy, the next factor 
of it originates from the zero-modes of 
the Dirichlet directions, and the 
second exponential specifies the dependence on 
the distance of the branes. Furthermore, the factor
$\prod_{n=1}^\infty(1- e^{-4n\pi t})^{p-d+3}$ is due to
the oscillators of the Dirichlet directions and the
conformal ghosts, while the second determinant
originates from the oscillators of the Neumann
directions. We observe that the interaction amplitude 
is exponentially damped by the square distance 
of the branes. Note that 
analogous analysis in the presence of an 
additional background field (i.e. the tachyon field) has been
worked out in Ref. \cite{23}.
Beside, similar results for a setup without rotation 
have been found in Ref. \cite{24}. For more investigation 
also see Refs. \cite{15}-\cite{20}.

\section{LREE corresponding to a D$p$-brane}

\subsection{Entanglement entropy of a bipartite system}

Let $|\psi \rangle$ denote the pure state
of the whole composite system, including the subsystems
A and B. The density operator which is associated to
this state is specified by $\rho=|\psi \rangle \langle \psi |$.
It satisfies
the probability conservation condition ${\rm Tr}\rho=1$.
Moreover, the reduced density matrix for the subsystem
A is defined by taking the partial trace over the
subsystem B as $\rho_{\rm A}={\rm Tr_B} \rho$.

Among the various quantities for measuring entanglement, the
entanglement entropy and the R\'{e}nyi entropy are more
interesting and attractive. The entanglement entropy is given
by the von Neumann formula
$S=- {\rm Tr} \left( \rho_{\rm A}\ln \rho_{\rm A}\right)$
\cite{25} and the R\'{e}nyi entropy is defined as
$S_n=\frac{1}{1-n} \ln {\rm Tr} \rho_{\rm A}^n$ with
$n\geq 0 ,\;n\neq1$ \cite{26}.
Note that the limit $n\rightarrow 1$
of the R\'{e}nyi entropy gives the
entanglement entropy.

\subsection{The density operator of the setup}

By expanding the exponential part of the state
(\ref{eq:2.6}) we receive a series which
elaborates an entanglement between the left-
and right-moving parts of the Hilbert space.
Since in our configuration all elements of the matrix
$S_{\mu\nu}$ are nonzero we have an extremely non-trivial
composite system with the left-right entanglement.

For a given boundary state
$|B\rangle$, associated with a D$p$-brane,
we may immediately take the density
matrix as $\rho=|B\rangle \langle B|$.
Since the inner product $\langle B|B \rangle$
is divergent, see Eq. (\ref{eq:3.2}) in the limit
$\epsilon \to 0$, this choice does not satisfy
the condition ${\rm Tr} \rho=1$.
Thus, a finite correlation length $\epsilon$
is introduced and the density matrix
is defined by \cite{27,28},
\begin{equation}
\rho=\frac{e^{-\epsilon H}|B\rangle
\langle B|e^{-\epsilon H}}{Z(2\epsilon)}~,
\end{equation}
where
$Z(2\epsilon)$ is fixed by ${\rm Tr} \rho =1$.
Therefore, the amplitude (\ref{eq:2.11}) conveniently
enables us to extract $Z(2\epsilon)$ as in the following
\begin{eqnarray}
Z(2\epsilon)&=&\langle B|e^{-2\epsilon H}|B\rangle
\nonumber\\[10pt]
&=&\frac{T_p^2\;V_{p+1}}{8(2\pi)^{d-p-1}}
|\det M |\bigg{[}e^{(d-2)\pi \epsilon/3}
\left(\sqrt{\frac{1}{4\epsilon}}\right)^{d-p-1}
~\nonumber\\[10pt]
&\times & \prod_{n=1}^\infty \bigg{(}
\det[\mathbf{1}-Q^{\rm T} Q e^{-8n\pi \epsilon}]^{-1}~
\left(1- e^{-8n\pi \epsilon}\right)^{p-d+3}\bigg{)}\bigg{]}
\label{eq:3.2} ~.
\end{eqnarray}
Note that the two interacting
boundary states exactly are alike, and their
corresponding branes
have been located at the same position. Consequently, the
$y$-dependent exponential disappeared
and also the indices 1 and 2 were omitted.
Hence, $Z(2\epsilon)$ can be manifestly interpreted as
the tree-level amplitude which a closed string
propagates for the time $2\epsilon$ between the 
very near D$p$-branes.

At first, we shall construct the LREE
corresponding to a bare-static brane
as a simple system, and then LREE
will be computed for a rotating-moving brane in the presence
of the Kalb-Ramond field and $U(1)$ gauge potential.

\subsection{LREE corresponding to a bare-static brane}

For this setup, quench the internal and background
fields, and also stop the rotation and linear
motion of the brane. Therefore, the partition
function (\ref{eq:3.2}) is simplified with
$\det M=-1$ and $Q^{\rm T}Q={\bf 1}$. In this case
we call it $Z_{(0)}(2\epsilon)$.
For deriving the R\'{e}nyi
entropy we need to compute  ${\rm Tr} \rho^n_{\rm L}$ for
the real number $n$, where the
subsystem ``L'' is the left-moving
part of the Hilbert space.
The replica trick enables us to accurately calculate
${\rm Tr} \rho^n_{\rm L}$, which yields
\begin{equation}
{\rm Tr} \rho_{\rm L}^n \sim \frac{Z_{(0)}(2n\epsilon)}
{Z_{(0)}^n (2 \epsilon )}
\equiv {\frac{Z_{(0)n}({\rm L})}{Z_{(0)}^n}}~,
\end{equation}
where $Z_{(0)n}({\rm L})$ is called ``replicated partition
function''.
By defining $q=e^{-4\pi \epsilon}$,
the last relation can be expressed in terms of the Dedekind
$\eta$-function
\begin{eqnarray}
\eta(q)=q^{1/12}\prod^\infty_{m=1}
\left(1-q^{2m}\right).
\nonumber
\end{eqnarray}
Since in the limit $\epsilon
\rightarrow 0$ the variable $q$ does not vanish,
the open/closed worldsheet
duality is employed to go to the open string
channel. Using the transformation
$4\epsilon \rightarrow 1/4\epsilon$
we obtain the new variable
$\tilde q=\exp {\left(-\frac{\pi}{4\epsilon}\right)}$
which vanishes at the limit $\epsilon \rightarrow 0$.
Hence, by expanding the
Dedekind $\eta$-function for small $\tilde{q}$, we acquire
\begin{eqnarray}
\frac{Z_{(0)n} ({\rm L})}{Z_{(0)}^n} &\approx &
K_0^{1-n}\left( \left( 2\sqrt{\epsilon}\right)^{1-n}
\sqrt{n}\right)^{d-p-1}\;
\exp{\left[ \frac{(d-2)\pi}{48\epsilon}
\left(\frac{1}{n}-n\right) \right]}
\nonumber\\[10pt]
&\times& \prod_{m=1}^{\infty} \bigg{\{}1+(d-2)\bigg{[}-n\;
e^{-{m\pi}/{2\epsilon}}+ \;e^{-{m\pi}/{2\epsilon n}}
\nonumber\\[10pt]
&-& n(d-2) \;e^{-(1+1/n){m\pi}/{2\epsilon}}
+\frac{d-1}{2}\; e^{-{m\pi}/{\epsilon n}}
\nonumber\\[10pt]
&+& \frac{n^2}{2}\left(d-2-\frac{1}{n}\right) \;
e^{-{m\pi}/{\epsilon}}\bigg{]}\bigg{\}}~,
\end{eqnarray}
where $K_0=T_p^2 V_{p+1}/8(2\pi)^{d-p-1}$.

Finally, by taking the limit $n \rightarrow 1$ of the
R\'{e}nyi entropy we receive the entanglement entropy
as in the following
\begin{eqnarray}
S_{(0){\rm LREE}}&=&\lim_{n \to 1}~\left[ \frac{1}{1-n}
\ln {\frac{Z_{(0)n}({\rm L})}{Z_{(0)}^n}} \right]
\nonumber\\[10pt]
&\approx &\ln K_0 + \frac{d-p-1}{2}\;\left(2 \ln {2}
+\ln \epsilon -1\right)
\nonumber\\[10pt]
&+&(d-2)\bigg{[} \frac{\pi}{24 \epsilon}
+  \left( 1-\frac{\pi}{2\epsilon}\right)\;
e^{-{\pi}/{2\epsilon}}+\frac{3}{2}
\left( 1-\frac{\pi}{\epsilon}\right)\;
e^{-{\pi}/{\epsilon}}\bigg{]}\label{eq:3.5}~,
\end{eqnarray}
up to the order
${\cal{O}}\left(\exp{(-3{\pi}/{2\epsilon})}\right)$.
The first term, i.e. $\ln{K_0}$, depends on the tension
and the worldvolume of the brane. It is related to the
boundary entropy of the brane.
In Refs. \cite{29, 30}
similar relations concerning the boundary entropy have been found.
However, the second factor denotes the zero-mode contribution
which originates from the Dirichlet directions.
The other terms are due to the oscillators. The factor
$-2$ in $(d-2)$ comes from the conformal ghosts.
The divergence $(d-2)\pi/24\epsilon$ can be justified by the
sum over all oscillating modes $\tilde \alpha_n$
which become more and more energetic \cite{11}.
For the special case $d=3$ and $p=1$,
the leading terms (the terms without exponential factors)
of Eq. (\ref{eq:3.5}) are exactly compatible with the result
of the Ref. \cite{11}.

\subsection{LREE corresponding to a dressed-dynamical brane}

Now we calculate the LREE regarding a
generalized configuration. Therefore,
our D$p$-brane possesses a tangential
dynamics and is dressed by the
background field $B_{\mu \nu}$ and the gauge
potential $A_\alpha$. In the previous section we obtained
the corresponding boundary state and the associated
partition function, i.e. Eq. (\ref{eq:3.2}).
Writing the ratio
${Z_n}/{Z^n}$ in terms of the Dedekind $\eta$-function,
and applying the transformation
$4\epsilon \to {1}/{4\epsilon}$ for receiving
the open string channel, and finally expanding the
$\eta$-function for small $\tilde{q}$,
give rise to the equation 
\begin{eqnarray}
\frac{Z_n}{Z^n}&\approx &  K^{1-n}
\left( \left( 2\sqrt{\epsilon}\right)^{1-n}
\sqrt{n}\right)^{d-p-1}\;
\exp{\left[ \frac{(d-2)\pi}{48\epsilon}
\left(\frac{1}{n}-n \right)\right]}
\nonumber\\
&\times& \prod_{m=1}^{\infty}\bigg{\{}1+(d-p-3)
\bigg{[}-ne^{-{m\pi}/{2\epsilon}}+\frac{n^2}{2}
\left(d-p-3-\frac{1}{n}\right) \;e^{-{m\pi}/{\epsilon}}
\nonumber\\[10pt]
&+& e^{-{m\pi}/{2\epsilon n}}-n(d-p-3)
e^{-(1+1/n){m\pi}/{2\epsilon}}
+\frac{(d-p-2)}{2}\; e^{-{m\pi}/{\epsilon n}}
\bigg{]}\bigg{\}}
\nonumber\\
&\times& \prod_{m=1}^{\infty}\bigg{\{} 1
-n\; {\rm Tr} (Q^{\rm T} Q)\;
e^{-{m\pi}/{2\epsilon}} -n \;[{\rm Tr} (Q^{\rm T} Q)]^2 \;
e^{-m\pi( {1}/{\epsilon}+{1}/{n\epsilon})}
\nonumber\\[10pt]
&+& {\rm Tr} (Q^{\rm T} Q) \;
e^{-{m\pi}/{2n\epsilon}}
+\frac{n}{2} \bigg{[}
-{\rm Tr} (Q^{\rm T} Q)^{2}
+ n [{\rm Tr} (Q^{\rm T} Q)]^2\bigg{]}e^{-{m\pi}/{\epsilon}}
\nonumber\\[10pt]
&+& \frac{1}{2}\bigg{[} {\rm Tr} (Q^{\rm T} Q)^2 \; +
[{\rm Tr} (Q^{\rm T} Q)]^2 \bigg{]}e^{-{m\pi}/{n\epsilon}}
\bigg{\}}~,
\end{eqnarray}
where $K=T_p^2 V_{p+1}|\det M |/8(2\pi)^{d-p-1}$.
The exponential in the first line
and the first infinite product are consequences of
the expansion of the $\eta$-function. Besides, the
second infinite product is due to the expansion
of the determinants in the partition function
and the replicated one.

The entanglement entropy of this generalized
configuration finds the feature
\begin{eqnarray}
S_{\rm LREE}&=&\lim_{n\to 1} \left[ \frac{1}{1-n}
\ln{\frac {Z_n}{Z^n}}\right]
\nonumber\\[10pt]
&\approx &\ln K + \frac{(d-p-1)}{2}\; \left( 2\ln {2}
+\ln {\epsilon}-1 \right)+\frac{(d-2)\pi}{24\epsilon}
\nonumber\\[10pt]
&+&\left( 1-\frac{\pi}{2\epsilon}\right)
\left[ d-p-3 + {\rm Tr} (Q^{\rm T} Q)\right]
e^{-{\pi}/{2\epsilon}}
\nonumber\\[10pt]
&+&\left( 1-\frac{\pi}{\epsilon}\right)
\left[ \frac{3}{2}(d-p-3) + {\rm Tr} (Q^{\rm T} Q)
+\frac{1}{2}{\rm Tr} (Q^{\rm T} Q)^2 \right]
e^{-{\pi}/{\epsilon}} \label{eq:3.7}~,
\end{eqnarray}
up to the order
${\cal{O}}\left(\exp{(-3{\pi}/{2\epsilon})}\right)$.
As it can be seen, the second and third phrases are the same
as for the bare-static D$p$-brane. In addition, the
effects of the background and internal fields
and the brane dynamics have been prominently accumulated
in the $Q$-dependent terms and $\ln K$.
However, by turning off the fields and stopping the
brane, Eq. (\ref{eq:3.7}) is reduced to the entanglement
entropy of the bare-static brane, as expected.

\subsection{Comparison with a thermal entropy}

We can associate the LREE of the dressed-dynamical brane,
i.e. Eq. (\ref{eq:3.7}), to the thermodynamics.
This resemblance can be done by defining a temperature
which is proportional to the inverse of the
infinitesimal parameter $\epsilon$. From this
point of view, the limit $\epsilon \to 0$ is
equivalent to the high temperature limit of the
thermal system. According to the partition function
(\ref{eq:3.2}), the thermodynamical entropy of the system
in the limit $\beta =2\epsilon \to 0$ takes the form
\begin{eqnarray}
S_{\rm th}&=&\beta^2 \frac{\partial}{\partial \beta}
\left( -\frac{1}{\beta} \ln {Z} \right)
\nonumber\\[10pt]
&\approx &\ln K + \frac{(d-p-1)}{2}\; \left[ 2\ln {2}
+\ln {\frac{\beta}{2}}-1 \right]
+\frac{(d-2)\pi}{12\beta}
\nonumber\\[10pt]
&+&\left( 1-\frac{\pi}{\beta}\right)
\left[ d-p-3 + {\rm Tr} (Q^{\rm T} Q)\right]
e^{-{\pi}/{\beta}}
\nonumber\\[10pt]
&+&\left( 1-\frac{2\pi}{\beta}\right)
\left[ \frac{3}{2}(d-p-3) + {\rm Tr} (Q^{\rm T} Q)
+\frac{1}{2}{\rm Tr} (Q^{\rm T} Q)^2 \right]
e^{-{2\pi}/{\beta}}~,
\end{eqnarray}
up to the order ${\cal{O}} \left(e^{-3\pi/\beta} \right)$.
We observe that this thermal entropy exactly is equal
to the LREE which was specified by Eq. (\ref{eq:3.7}).
In fact, these two entropies basically are
different quantities. This desirable connection
may reveal a close relation between
the entanglement entropy and thermodynamic entropy.
There are also other works which illustrate such 
connections.
For instance, the Refs. \cite{31,32,33} provide a relation 
similar to the first law of thermodynamics via the 
entanglement entropy. 

\section{Some simple configurations with $p=2$}

For clarifying our results,
we reduce the general complicated case to
the D2-brane. Let the brane sit on the $x^1x^2$-plane.
The matrices for the D2-brane are given by
\begin{equation}
\omega_{\alpha \beta}=
\begin{pmatrix}
0 & v_1 & v_2\\
-v_1 & 0 & \Omega\\
-v_2 & -\Omega & 0\\
\end{pmatrix}
\quad~,~ ~ \mathcal{F}_{\alpha \beta}=
\begin{pmatrix}
0 & \mathcal{E}_1 & \mathcal{E}_2 \\
-\mathcal{E}_1 & 0 & \mathcal{B} \\
-\mathcal{E}_2 & -\mathcal{B} & 0 \\
\end{pmatrix},
\end{equation}
where the parameters exhibit the following quantities,

$v_1$ and $v_2$: the components of the linear velocity
of the brane,

$\Omega$: the angular velocity of the brane,

$\mathcal E_1$ and $\mathcal E_2$: the components of
the total electric field inside the brane,

${\mathcal B}$: total magnetic field,

in which $\mathcal E_1 \equiv \mathcal F_{0 1}
=F_{0 1}-B_{0 1}$ (similarly for  $\mathcal E_2$),
$\Omega = \omega_{12}$
and $\mathcal B\equiv \mathcal F_{1 2} =F_{1 2}
-B_{1 2}$. For more illustration 
we shall decompose this setup
into the following special configurations.

{\bf A dressed-boosted D2-brane}

At first let us only turn on $v_1$ and $\mathcal{E}_2$.
In fact, to have a sensible tangential velocity
$v_1$, presence of the electric field
$\mathcal{E}_2$ is inevitable,
otherwise, the Lorentz invariance is restored
and consequently there is no
preferable direction. For this case we obtain
\begin{equation}
{\rm Tr} \left(Q^{\rm T} Q\right)=3+\;
\frac{4v^2_1{\mathcal{E}^2_2}}{1-v^2_1-{\mathcal{E}^2_2}}~.
\end{equation}
Hence, the LREE (up to the order $e^{-\pi/\epsilon}$)
becomes
\begin{eqnarray}
S_{(1)\rm LREE}
\approx S_0+\frac{4v^2_1\;\mathcal{E}^2_2}{1-v^2_1
-\mathcal{E}^2_2}\;\left(1-\frac{\pi}{2\epsilon}\right)\;
e^{-\pi/2\epsilon}~,
\end{eqnarray}
where $S_0$ denotes the LREE for a bare-static D2-brane.
It is given by Eq. (\ref{eq:3.5}) with $p=2$.

The prefactor of Eq. (\ref{eq:2.6}) gives rise to
the condition $\det M <0$. On the basis of this, 
the denominator
of Eq. (4.3) for a moving D2-brane does not vanish,
i.e. $1-v^2_1 -\mathcal{E}^2_2 > 15 v^2_1$.
Thus, the entropy $S_{(1)\rm LREE}$ for any finite value of
$\mathcal{E}_2$ satisfactorily remains finite.

As the second special case, we consider $v_1$ and
$\mathcal{B}$ to be nonzero. Hence, the trace factor is given by
\begin{equation}
{\rm Tr}\left(Q^{\rm T} Q\right)=3-\frac{4v^2_1{\mathcal{B}}^2}
{1-v^2_1+{\mathcal{B}}^2}~.
\end{equation}
Therefore, the LREE takes the form
\begin{eqnarray}
S_{(2)\rm LREE}
\approx S_0-\frac{4v^2_1{\mathcal{B}}^2}{1-v^2_1
+{\mathcal{B}}^2}\;\left(1-\frac{\pi}{2\epsilon}\right)\;
e^{-\pi/2\epsilon}~.
\end{eqnarray}

{\bf A dressed-rotating D2-brane}

Another profitable option is illustrated by turning on the
fields and the brane rotation.
Again note that the fields are indeed necessary
for sensibility of the tangential rotation.
At first, consider a
rotating D2-brane with the angular velocity $\Omega$
which is dressed with $\mathcal{E}_1$.
Accordingly, we receive the following LREE
\begin{eqnarray}
S_{(3)\rm LREE}
\approx S_0-\frac{4{\Omega}^2{\mathcal{E}^2_1}}
{1+{\Omega}^2-{\mathcal{E}^2_1}}\;
\left(1-\frac{\pi}{2\epsilon}\right)\;
e^{-\pi/2\epsilon}~.
\end{eqnarray}

For the last case, we turn on the angular velocity
$\Omega$ and the magnetic
field $\mathcal{B}$, which yield
\begin{eqnarray}
S_{(4)\rm LREE} \approx S_0+\; \frac{8 \Omega
\mathcal{B}}{1+(\Omega-\mathcal{B})^2}\;
\left(1-\frac{\pi}{2\epsilon}\right)\;e^{-\pi/2\epsilon}~.
\end{eqnarray}

Note that similar to the finiteness of $S_{(1)\rm LREE}$,
again the condition $\det M <0$ eventuates to the
finiteness of the other three foregoing entropies.

\section{Conclusions}

At first, we acquired the LREE of a bare-static D$p$-brane.
Then, the LREE of a rotating-moving D$p$-brane in the presence
of the Kalb-Ramond background field and an internal
$U(1)$ gauge potential was computed. For this purpose, we utilized
the boundary state, associated with the D$p$-brane,
and the interaction amplitude between the two
identical and parallel D$p$-branes.
For the dressed-dynamical brane
presence of the various parameters
in the setup dedicated a generalized feature to the LREE.
By varying the parameters the value of the
LREE can be accurately adjusted to any desirable value.

The partition function enabled us to conveniently 
calculate a reliable thermodynamic entropy. 
The LREE of the dressed-dynamical D$p$-brane
was compared with this 
entropy. We observed that, by redefinition
of the temperature, the two entropies
exactly are the same. This connection may be useful  
for the future works. For example, by deriving the 
LREE for those supersymmetric configurations 
which represent the black holes, 
one may find the Bekenstein-Hawking entropy.

Finally, for explicit appearance of the various parameters,
we reduced the general case to the D2-brane
with either a linear velocity or an
angular velocity in the presence of the total
electric field $\mathcal{E}$
or the total magnetic field $\mathcal{B}$.


\end{document}